\documentclass{article}
\usepackage{graphicx} 
\usepackage{amsmath, amssymb}
\usepackage{hyperref}
\usepackage[top=1in,bottom=1in,left=1in,right=1in]{geometry}

\title{On the dynamical evolution of randomness\\Part A: Random experiments as dynamical systems}
\author{Allen Lobo\thanks{Dept. of Physics, St. Joseph's University. e-mail: allen.e.lobo@outlook.com} \ \& Saravanan Arumugam\thanks{Department of Physics, Sikkim Manipal Institute of Technology, Sikkim Manipal University. e-mail: sarvaanphysics@gmail.com}}
\date{February 2025}

\begin{document}

\maketitle
\begin{abstract}
    In this work, Bernoulli's Law of Large Numbers, also known as the Golden theorem, has been extended to study the relations between empirical probability and empirical randomness of an otherwise random experiment. Using the example of a coin toss and a dice role, some interesting results are drawn. Analytically and using numerical computations, empirical randomness of each outcome has been shown to increase by \textit{ chance}, which itself depends on the growth rate of empirical probabilities. The analyses presented in this work, apart form depicting the nature of flow of random experiments in repetitions, also present dynamical behaviours of the random experiment, and experimental and simulation-based verifications of the mathematical analyses. It also presents an appreciation of the beauty of Bernoulli's Golden theorem and its applications by extension.
\end{abstract}
\section{Introduction}

Historically, in as early as the $16^{th}$ century, Gerolamo Cardano (in his book \textit{Liber de ludo aleae}) \cite{Bellhouse2005DecodingAleae} observed a growth of the accuracy of his observations when the number of repetitions were increased. This was followed in the $18^{th}$ century when Bernoulli presented a mathematics behind it (in his work, \textit{Ars Conjectandi}) \cite{Mattmuller2014The1713} as the Law of Large Numbers \cite{Bolthausen2013BernoullisNumbers}.

Since then, the Law of Large Numbers has presented an intriguing aspect of the classical probability theory. The Law of Large Numbers (LLN), or Bernoulli's Golden theorem (as termed by Bernoulli) presents a relation between theoretically defined probability of outcomes of an experiment and the statistics of the outcomes. It states that when the experiment is repeated for a \textit{large number} of times, the empirically observed statistics of the outcomes match the theoretically predicted probabilities. 

\maketitle
Since its mathematical formulation by Bernoulli, the LLN has since been presented and modified by many authors \cite{Chibisov2016BernoullisNumbers, Teran2008OnNumbers, Weba2009AMethod, Goldstein1975SomeNumbers, Dedecker2007ApplicationsNumbers, Teran2006AApplications, SHIRIKYAN2003AAPPLICATIONS, Yang2008AApplications, Kay2014LawsApplications}, such as the Strong Law of Large Numbers presented by Kolmogorov and the weak law presented by Khinchin. 

However, the Golden theorem can be analyzed further in order to assess some more aspects of the probabilistic nature of random experiments. One such aspect is the idea of the degree of "randomness" of an experiment. Randomness, in itself, is a condition of unpredictability of the outcomes of an experiment. Although this necessarily dictates that each outcome of an experiment is equally unbiased, the influence of \textit{chance} can significantly deviate the empirical probability from the theoretical one. This occurs when, upon successive trials, one output seems to be more biased than others. The LLN dictates that as the number of trials tend towards infinity, this bias reduces to zero, and the empirical probability of each outcome gradually converges to the theoretical predictions. This convergence to the theoretical values is thought to depend on chance, since over repeated experiments, each outcome occurs for a sufficient number of times as the number or repetitions tend towards infinity.

Suppose $a$ and $b$ are the outcomes of a random experiment, with equal probabilities $P$ of occurrence, such that 
\begin{equation}
P_a = P_b .
\end{equation}
As the number of trials $N$ approaches infinity, the number of outcomes of both $a$ $(=f_a)$ and $b$ $(=f_b)$ must increase.
\begin{gather}
f_a + f_b = N. \\
\because N\rightarrow \infty,\quad \therefore f_a + f_b \rightarrow\infty.
\intertext{This implies that one of the following three cases must be satisfied:}
f_a + f_b\rightarrow \infty \Leftrightarrow
\begin{cases}
    \text{Only } f_a \rightarrow \infty, \text{ or,}\\
    \text{Only }f_b \rightarrow \infty \text{ or,}\\
    \text{Both }f_a \text{ and } f_b \rightarrow \infty.
\end{cases}
\end{gather}
Both the first and second conditions imply a bias in the outcomes and are against the idea of random outcomes. It therefore makes mathematical sense to state that in a random experiment, for large repetitions, the empirical probabilities $P^\epsilon$ of each outcome, which are ratios of individual number of outcomes to total number of trials, become equal. Let $ P^\epsilon_a$ and $ P^\epsilon_b$ be the empirical probabilities of the outcomes $a$ and $b$, such that
\begin{gather}
    P^\epsilon_a = \frac{f_a}{N}, \quad P^\epsilon_b = \frac{f_b}{N}.
    \end{gather}
When $N\rightarrow\infty$, both $P^\epsilon_a$ and $P^\epsilon_b$ approach irrational forms of ${\infty}/{\infty}$. However, the condition
\begin{equation}
    P^\epsilon_a + P^\epsilon_b = \frac{f_a + f_b}{N} \equiv 1
\end{equation}
must be satisfied. For the same, a limit can be introduced as
\begin{equation}
    \lim_{N \to \infty} \quad f_b\rightarrow f_a.\label{infinity_cond_for_fi_fj}
\end{equation}
This directly points towards the solution that:
\begin{equation}
    P^\epsilon_a = P^\epsilon_b,
\end{equation}
which agrees with the theoretical case.

Hence, theoretical probability can be safely regarded as a simple measurement of the possibilities of outcomes in an \textit{ideal} scenario, which in no way dictates the occurrence of those outcomes. This idea about theoretical probability, although well-understood, is almost counter-intuitive to the reader who, in the case of a random experiment, expects outcomes to occur in the (theoretically) predicted fashion. This therefore presents the following questions:
\begin{enumerate}
\item If, upon repetitions, one outcome occurs more frequently than the other(s), is the experiment truly random? In other words, \textbf{\textit{how empirically random} is the random experiment}?
    \item If each outcome bears an equal probability of occurrence in an ideal random experiment, what dictates the outcomes in each repetition? In other words,\textbf{ what is the mathematics behind \textit{chance}}?
\end{enumerate}
 The brief work presented in this communication aims to present arguments and mathematics which aim at answering the enumerated questions above. In order to test the presented mathematical analyses, numerical simulations and quantum experiments have been conducted using computational set-ups. Specifically for the quantum experiment, we have used the IBM-quantum computational facility. We then explore the dynamics of the random experiments and present an explanation for the occurrence of the Law of Large Numbers, based on the observations.
 
\section{Mathematical Analyses}
 We consider a random experiment with $m$ outcomes, each having a random theoretical probability $P_i$ (which is \textit{ideally} equal to $1/m$ to represent a purely unbiased set-up), where $i\in [1,m]$. We repeat the experiment $n$ number of times, and measure the frequency of each outcome as $f_i$. 

The empirical probability $P_i^\epsilon$ is then calculated as the ratio of the outcome frequency to total number of repetitions, 
\begin{equation}
    P_i^\epsilon = \frac{f_i}{n}.
\end{equation}

In this communication, a new measure of the "randomness" of a system has been presented. Since, classically, randomness is measured using the entropy $(\sigma)$ of the system, we define our measurement of randomness as the $\Lambda-$entropy $(\Lambda_\sigma)$. We define $\Lambda-$entropy as the joint empirical probability of all outcomes,
\begin{equation}
    \Lambda_\sigma = \prod_{i=1}^{m}P^\epsilon_i = \prod_{i=1}^m\frac{f_i}{n} = \frac{1}{n^m}\prod_{i=1}^m f_i.
\end{equation}

The golden theorem dictates that the empirical probabilities must converge to the theoretical probabilities as $n\rightarrow\infty$. Hence, the $\Lambda-$entropy must also converge to a point $\Lambda_\sigma^0$, where,
\begin{equation}\label{lambda_sigma_eqn}
    \Lambda_\sigma^0 = \prod_{i=1}^m P_i.
\end{equation}
Hence, at-least two approximations can be presented here in regard to the $\Lambda-$entropy as per Bernoulli's LLN, which states that
    as the number of trials (repetitions) of the random experiment increase towards infinity, according to the Law of Large Numbers, the $\Lambda-$entropy saturates to $\Lambda_\sigma^0$, i.e.,
    \begin{gather}
         \lim_{n\to\infty}\Lambda_\sigma = \Lambda_\sigma^0,\label{axiom1.1}\\
         \lim_{n\to\infty}\frac{d}{dn}\Lambda_\sigma =0.\label{axiom1.2}
    \end{gather}

It can be readily realised that for an ideal random experiment, that is to say, for an experiment with no bias in its outcomes, the $\Lambda-$entropy should be equal to $\Lambda_\sigma^0$. Hence, as the experiment starts and proceeds with repetitions such that $n\geq m$, $\Lambda_\sigma$ may deviate away from $\Lambda_\sigma^0$. This will occur not due to bias in the outcomes (since the experiment is an ideal random experiment) but due to \textit{chance}. Hence, the empirical randomness of the experiment can be measured by measuring the relative shift of $\Lambda_\sigma$ from the $\Lambda_\sigma^0$ value. This approach provides an interesting measure of the empirical randomness of a random experiment. Alternatively speaking, if an experiment with well-defined outcomes is more random, as the experiment proceeds into repetitions, the $\Lambda-$entropy remains closer to its ideal value $\Lambda_\sigma^0$. The greater $\Lambda_\sigma$ shifts from the $\Lambda_\sigma^0$ value, lesser is the empirical randomness of the experiment. i.e.,
    \begin{equation}
       \text{Empirical Randomness} \propto \frac{1}{|\Lambda_\sigma - \Lambda_\sigma^0|}.\label{theorem1.1}
    \end{equation}
    This deviation of the empirical randomness as $\Lambda_\sigma$ deviates from $\Lambda_\sigma^0$ can be measured using a parameter $\sigma_\epsilon$, where,
    \begin{gather}
        \sigma_\epsilon = |\Lambda_\sigma - \Lambda_\sigma^0|,
        \intertext{such that,}
        \text{Empirical Randomness} \propto \frac{1}{\sigma_\epsilon}.
    \end{gather}

We term $\sigma_\epsilon$ as the empirical randomness parameter. Empirical randomness increases as $\sigma_\epsilon$ approaches zero. Equation (\ref{theorem1.1}) provide a measure of the true empirical randomness of an otherwise random experiment. It states that even though, theoretically, the experiment presents outcomes randomly in each trial, the outcomes occur in some bias presented by \textit{chance} and as such are not truly empirically random. As the outcomes occur in such a way that no outcome seems to occur with bias, which is to say that no outcome appears to occur more \textit{favourably} or \textit{luckily} than others, the outcomes reach a true state of randomness, which is marked by the convergence of the $\Lambda-$entropy towards $\Lambda_\sigma^0$. 

Upon attempting to investigate how the $\Lambda-$entropy varies with the trials of the experiment, we find its derivatives with respect to $n$. Here, it is important to mention that over a large number of repetitions, $n$ and $\Lambda_\sigma$ can be treated as continuous variables. The validity of these treatments can be verified in the numerical experiments which we present later in this communication.

Differentiating $\Lambda_\sigma$ with respect to $n$, we get --
\begin{gather}\allowdisplaybreaks
    \frac{d\Lambda_\sigma}{dn} = \frac{d}{dn}\left(\frac{1}{n^m}\prod_{i=1}^{m}f_i\right) \\
    = \left(\prod_{i=1}^{m}f_i\right)\frac{d}{dn}\frac{1}{n^m} + \frac{1}{n^m}\left(\frac{d}{dn}\prod_{i=1}^{m}f_i\right).\\
    = -\frac{m}{n}\frac{1}{n^m}\left(\prod_{i=1}^{m}f_i\right) + \frac{1}{n^m}\sum_{i=1}^m\left(\prod_{j=1}^{m}f_j\right)\frac{1}{f_i}\frac{df_i}{dn}.\\
   \therefore \frac{d}{dn}\Lambda_\sigma = \Lambda_\sigma \left( -\frac{m}{n} + \sum_{i=1}^m \frac{d }{dn}\ln f_i \right). 
   \intertext{The summation in the above equation can be further reduced to simplify the form, giving --}
   \frac{d}{dn}\Lambda_\sigma = \Lambda_\sigma \left[ \frac{d }{dn}\ln \left(\prod_{i=1}^m f_i\right) -\frac{m}{n} \right].\label{derivative_of_lambda_entropy_first_principle}
\end{gather}

It can be seen from equation (\ref{derivative_of_lambda_entropy_first_principle}) that the number of outcomes $(m)$ reduces the convergence rate of $\Lambda-$entropy towards the $\Lambda_\sigma^0$ value. It may be noted that in the case where $n<m$, at-least one outcome has not yet occurred, such that both $\prod_{i=1}^m f_i$ and $\Lambda_\sigma$ remain zero. Hence, $\Lambda_\sigma$ gains magnitude only when all the outcomes have occurred at-least once. Alternatively, for the case when $m\rightarrow\infty$, it can be shown that $\Lambda_\sigma$ remains zero until the number of trials itself becomes large enough, such that $\Lambda_\sigma$ becomes equal to $\Lambda_\sigma^0$.

For small values of $n$, provided the condition that $n\geq m$, simple manipulation of equation (\ref{derivative_of_lambda_entropy_first_principle}) presents the nature of \textit{flow} of outcomes during the repetition of the experiment. From equation (\ref{derivative_of_lambda_entropy_first_principle}) we have,
\begin{gather}
    \frac{1}{\Lambda_\sigma}\frac{d}{dn}\Lambda_\sigma  +\frac{m}{n}= \frac{d }{dn}\ln \left(\prod_{i=1}^m f_i\right) = \left(\prod_{i=1}^m f_i^{-1}\right)  \frac{d }{dn}\prod_{i=1}^m f_i\\
    \Rightarrow  \frac{d }{dn}\prod_{i=1}^m f_i=\left(\prod_{i=1}^m f_i\right)\cdot\left[\frac{d}{dn}\left(\ln\Lambda_\sigma\right)  +\frac{m}{n} \right].
    \intertext{Expressing the above expression in terms of the empirical randomness parameter $(\sigma_\epsilon)$,}
    \begin{split}
    \frac{d }{dn}\prod_{i=1}^m f_i=\left(\prod_{i=1}^m f_i\right)&\cdot\left[\frac{d}{dn}\ln\left(\Lambda_\sigma^0 + \delta\sigma_\epsilon\right)  +\frac{m}{n} \right],\\
    \text{where,}\quad\delta &= \begin{cases}
        +1, & \Lambda_\sigma > \Lambda_\sigma^0\\
        -1, & \Lambda_\sigma < \Lambda_\sigma^0\\
        0, & \Lambda_\sigma = \Lambda_\sigma^0.
    \end{cases}.
    \end{split}\label{chance_equation}
\end{gather}

Equation (\ref{chance_equation}) states that the rate at which outcomes of an experiment occur such that each outcome gains its frequency of occurrence as the number of trials increase, depend directly on how empirically random the experiment is at its current stage, which is measured by $\sigma_\epsilon$. In other words, if the random experiment proceeds in such a way that the frequency of outcomes deviate from uniformity, the empirical randomness parameter $(\sigma_\epsilon)$ increases, due to which the rate at which $\prod_{i=1}^m f_i$ grows per trial also increases, which implies that outcomes with currently less occurrence frequencies $f_i$ also increase such that $\prod_{i=1}^m f_i$ grows more rapidly. Outcomes of a random experiment are therefore not merely governed by \textit{chance}, but depend on the instantaneous empirical randomness of the experiment as it is repeated. Indeed as $n\rightarrow \infty$, equation (\ref{chance_equation}) dictates that the product $\prod_{i=1}^m$ reaches a maxima.

In order to further explore this relation between the rate of occurrence of each outcome, we investigate into equation (\ref{derivative_of_lambda_entropy_first_principle}) and determine the evolution of the difference of the frequencies of each outcome. We initially take the case of an ideal experiment with two outcomes $(m=2)$, such as the coin-toss experiment. From equation (\ref{derivative_of_lambda_entropy_first_principle}), we get:
\begin{gather}
    \frac{d}{dn}\ln{\Lambda_\sigma} + \frac{m}{n} = \sum_{i=1}^{m} \frac{1}{f_i}\frac{d}{dn}f_i = \frac{1}{f_H}\frac{df_H}{dn}+\frac{1}{f_T}\frac{df_T}{dn},
    \intertext{where $f_H$ and $f_T$ represent the frequencies of occurrence of Head and Tail as outcomes of the coin toss, respectively. Multiplying both sides by $n$ gives:}
    n\frac{d}{dn}\ln{\Lambda_\sigma} + m =  \frac{n}{f_H}\frac{df_H}{dn}+\frac{n}{f_T}\frac{df_T}{dn}=\frac{1}{P^\epsilon_H}\frac{df_H}{dn}+\frac{1}{P^\epsilon_T}\frac{df_T}{dn}\\
    \Rightarrow \frac{df_H}{dn} = -\frac{P^\epsilon_H}{P^\epsilon_T}\frac{df_T}{dn} + mP^\epsilon_H + nP^\epsilon_H\frac{d}{dn}\ln{\Lambda_\sigma}\\
    \Rightarrow \frac{df_H}{dn} = -\frac{P_H^\epsilon}{P_T^\epsilon}\frac{df_T}{dn} + mP^\epsilon_H + f_H\frac{d}{dn}\ln{\Lambda_\sigma}.\\
    \therefore\frac{d}{dn}(f_H - f_T) = \frac{df_H}{dn} - \frac{df_T}{dn} = mP^\epsilon_H + f_H\frac{d}{dn}\ln{\Lambda_\sigma}- \frac{df_T}{dn}\left( 1+\frac{P_H^\epsilon}{P_T^\epsilon}\right).
\end{gather}
 The above expressions can be now generalised as follows:
 \begin{gather}
     \frac{df_j}{dn} = f_j\frac{d}{dn}\ln{\Lambda_\sigma} + P^\epsilon_j m - \sum_{j\neq i}^{m}\frac{P_j^\epsilon}{P_i^\epsilon}\frac{df_i}{dn},\label{gen_eqn1}\\
     \frac{d}{dn}\left(f_j-f_i\right)_{\forall i\neq j} = f_j\frac{d}{dn}\ln{\Lambda_\sigma} + P^\epsilon_j m - \frac{df_i}{dn} - \sum_{j\neq i}^{m}\frac{P_j^\epsilon}{P_i^\epsilon}\frac{df_i}{dn}.\label{gen_eqn2}
 \end{gather}
 From both equations (\ref{gen_eqn1}) and (\ref{gen_eqn2}), the original equation (\ref{derivative_of_lambda_entropy_first_principle}) can be extracted by multiplying with $f_i^{-1}$ on both sides. These equations may be integrated accordingly to present an expression of $f_j(n)$ and its difference with other outcome frequencies:
 \begin{equation}\label{integration}
     f_j(n) = \int_0^n\left( f_j\frac{d}{dn}\ln{\Lambda_\sigma} + P^\epsilon_j m - \sum_{j\neq i}^{m}\frac{P_j^\epsilon}{P_i^\epsilon}\frac{df_i}{dn}\right) dn.
 \end{equation}
 Here, one must note that the equation holds validity only after each outcome has occurred at-least once, such that $P_i^\epsilon>0$, to avoid singularity. In the limit $n\rightarrow\infty$, the following adjustments can be made, as previously discussed:
 \begin{gather}
     \lim_{n\to\infty}\begin{cases}
         \Lambda_\sigma \rightarrow\Lambda_\sigma^0,\\
         \frac{d}{dn}\ln\Lambda_\sigma \rightarrow 0,\\
         P_i^\epsilon \rightarrow P_i=\frac{1}{m},\\
         \therefore\sum_{j\neq i}^{m}\frac{P_j^\epsilon}{P_i^\epsilon}\frac{df_i}{dn} = \sum_{j\neq i}^{m}\frac{df_i}{dn}= \frac{d \sum_{j\neq i}^{m}f_i}{dn} = 1,\\
         \text{and, }f_{i,j} \rightarrow\infty.
     \end{cases}
     \intertext{This modifies equation (\ref{gen_eqn1}) as follows:}
     \lim_{n\to \infty} \frac{df_j}{dn} = \mathcal{E} + 1 - 1 = \mathcal{E},\label{uncertainity_inlimit}
 \end{gather}
 where,  $\mathcal{E} = \lim_{n\to\infty}f_j\frac{d}{dn}\ln{\Lambda_\sigma} = \infty\times 0$ is the indeterminate term. What is interesting is that in this limit, the rate of growth of each outcome becomes equal to $\mathcal{E}$, independent of the growth rate of other outcomes, as is seen in the case of small values of $n$.The value of $\mathcal{E}$ can, however, only belong to the finite set $\{0,1\}$, since at each trial, the outcome frequencies can increase either by $1$ or remain same. This is valid even for the case of finite values of $n$. Also, only one outcome can have $\mathcal{E}=1$ at each trial, with the others having growth rates as $0$. The indeterminate form in this case arises by the inability of prediction of the growth rate of the outcomes based on the growths of other outcomes, as seen in equations (\ref{gen_eqn1}) and (\ref{gen_eqn2}). This means that in the limit $n\rightarrow \infty$, the outcomes of the experiment become truly indeterminable. They do not follow the probability evolution in the form of equations (\ref{gen_eqn1}) or (\ref{derivative_of_lambda_entropy_first_principle}), and therefore can not be predicted, such as in equation (\ref{integration}). Therefore, the outcomes achieve a true state of \textit{randomness}, which the theoretical model predicts. It therefore makes sense that in such a state, the experiment displays natures of ideal randomness, which is confirmed by the divergence of $1/\sigma_\epsilon$ to infinity and convergence of $\Lambda_\sigma$ to $\Lambda_\sigma^0$. This also implies that as the experiment proceeds, the term $f_i-f_j$ must obtain truly random values, since the result would depend on the integration of an indeterminable term:
 \begin{gather}\label{ft-fh}
   \forall i\neq j, \quad \lim_{n\to\infty}  f_i - f_j = \int\mathcal{E}dn  - f_j.
 \end{gather}
  
 The mathematics presented herein describes how the outcomes of a random experiment behave when the experiment itself is repeated. Specifically, we presented in this section a measurement of the empirical randomness of a random experiment. We showed that \textit{chance}, which is thought to decide which outcome will occur at each repetition of a random experiment, actually bears a mathematical relation. We showed that this relation can be used to determine the rate of occurrence of each outcomes during the successive trials of the random experiment. The analyses present answers to the previously enumerated questions which we listed in the first section of this communication. We proceed to the next section wherein we present some numerically simulated experiments to validate or analytical findings. We then present an experimental investigation of the accuracy of our analytical results using quantum-experimental setups. 
\section{Numerical Simulation and Experimental Verification of the analyses}
The mathematical development in the previous section can be employed to study the evolution of probability of outcomes of a random experiment. The calculations presented in equations (\ref{lambda_sigma_eqn}-\ref{uncertainity_inlimit}) can be then performed and the corresponding results be presented in order to discuss the accuracy and applicability of the mathematics. Alternatively, we investigate the applicability of our findings using quantum experiments performed using quantum computing facility. We present the results in this section.

We begin with the study of a single coin-toss experiment, in which a number out of the set $\{1,2\}$ is drawn at random using the \texttt{randi} function in \texttt{MATLAB}, where $1$ corresponds to head $(H)$ and $2$ corresponds to tail $(T)$. 

\begin{figure}[!ht]
    \centering
    \includegraphics[width=0.48\linewidth]{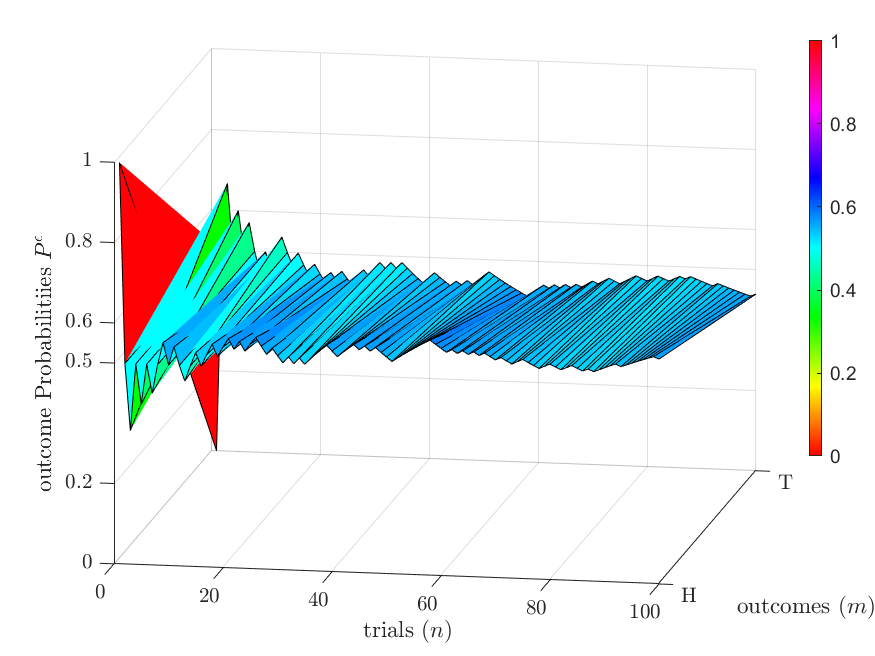}
    \includegraphics[width=0.48\linewidth]{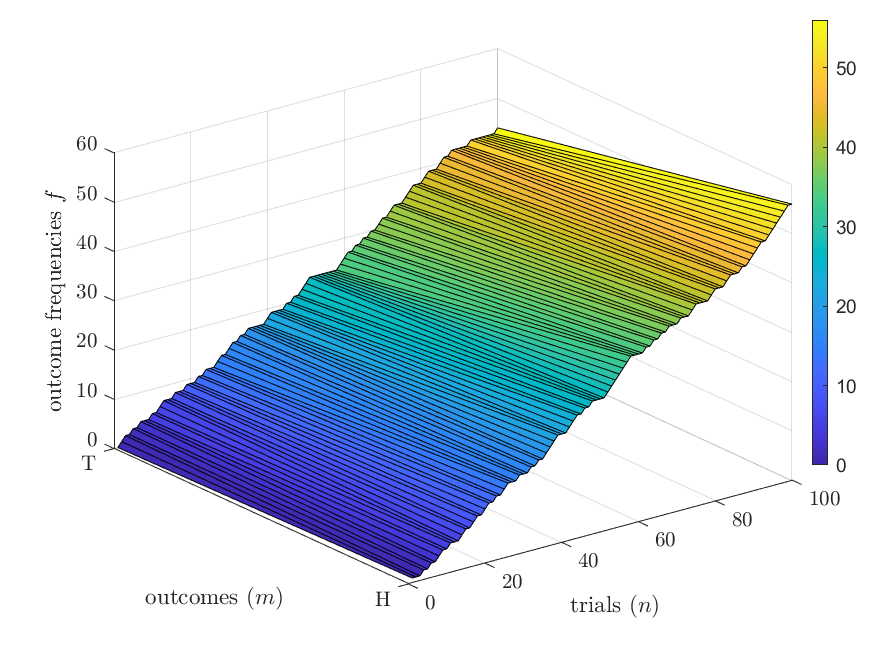}
    \includegraphics[width=0.5\linewidth]{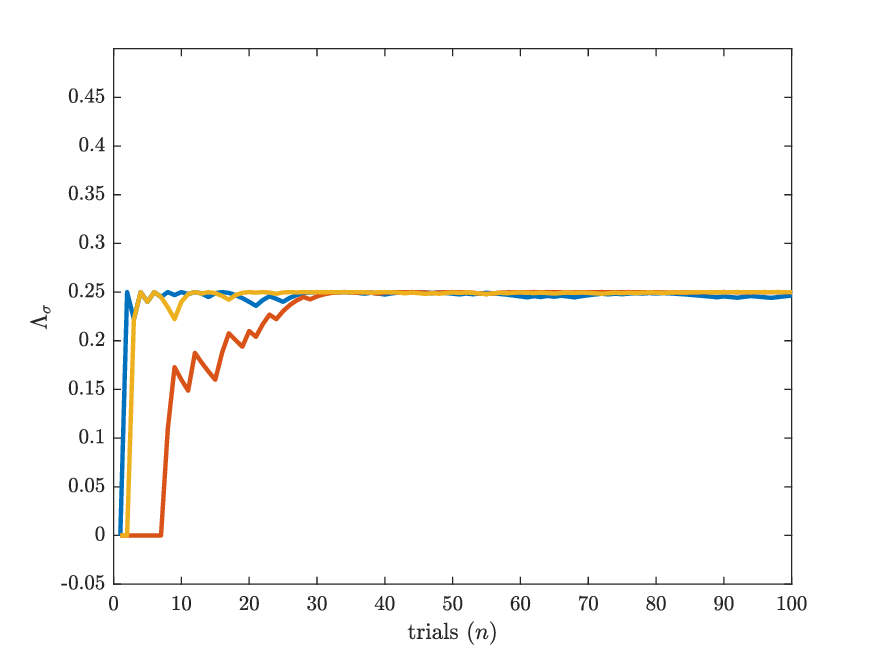}
    \caption{Evolution of (top, left) probabilities and (top, right) outcome frequencies in the single coin-toss simulation. (bottom) Evolution of the $\Lambda_\sigma$ entropy.}
    \label{fig:p_and_f_evolution}
\end{figure}

\begin{figure}[!ht]
    \centering
    \includegraphics[width=1\linewidth]{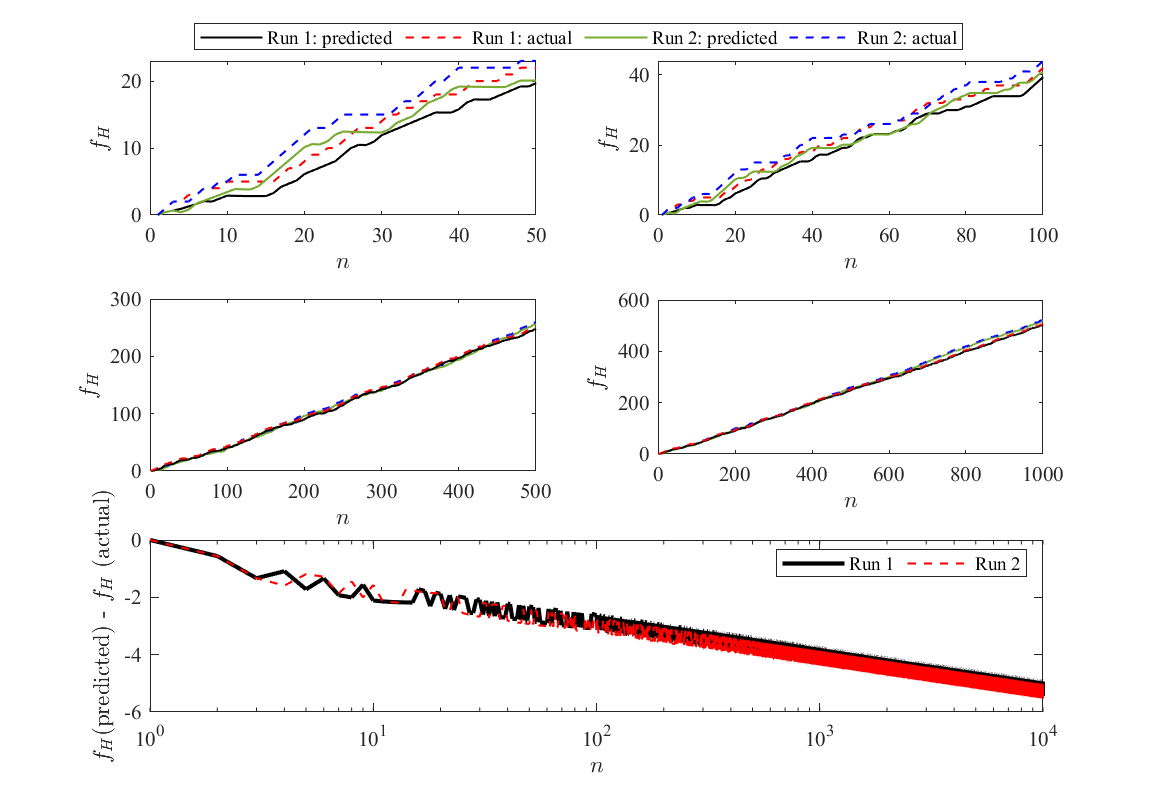}
    \caption{Evolution of outcomes: (from top, 1st and 2nd row) predicted growth of head outcome frequencies compared with actual growths of $f_H$, shown for various trials. (bottom row) Growth of uncertainty as the number of trials increase, as predicted by equation (\ref{uncertainity_inlimit}).}
    \label{fig:f_H_evols}
\end{figure}

\begin{figure}[!ht]
    \centering
    \includegraphics[width=0.9\linewidth]{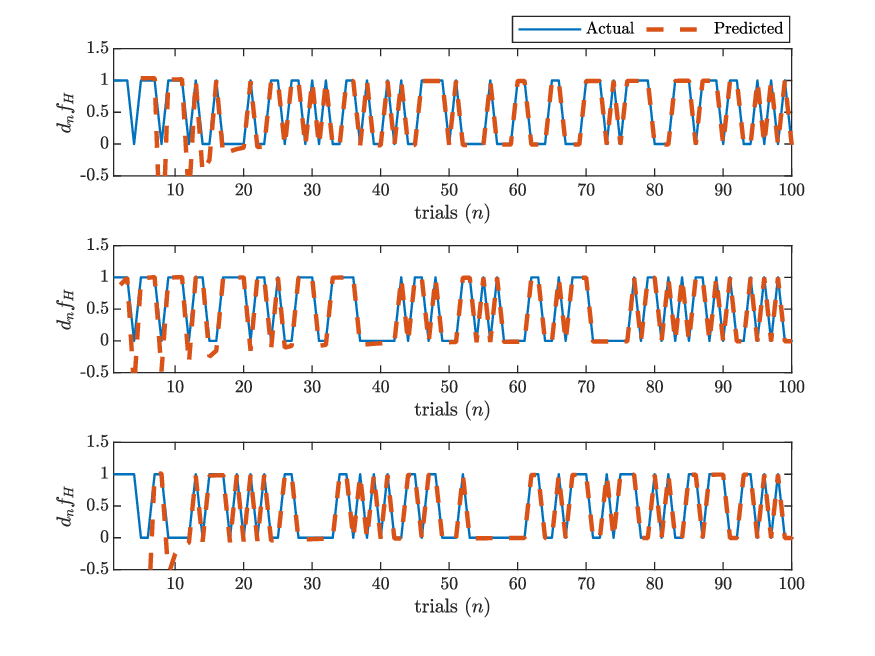}
    \caption{Growth rate of frequency of occurrence of Head: predicted growth rate using equation (\ref{gen_eqn1}) compared to actual growth rates, shown for different runs of the simulation (different rows).}
    \label{fig:f_gr_ct_S}
\end{figure}

\begin{figure}[!ht]
    \centering
   \includegraphics[width=0.8\linewidth]{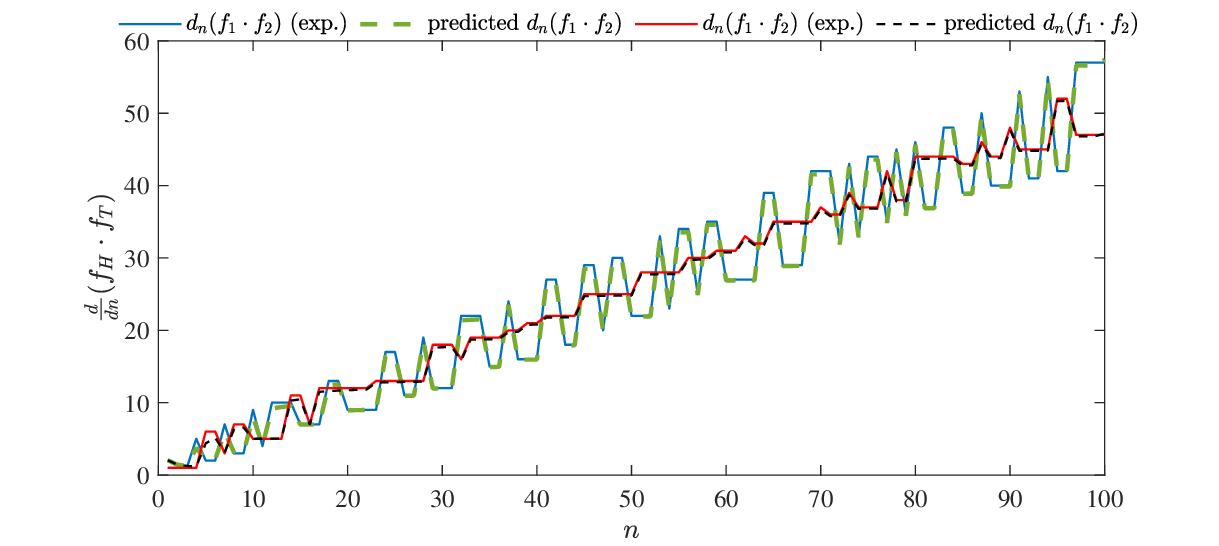}
    \caption{Growth-rates of the product of frequencies $f_H\cdot f_T$ shown for two simulation runs, comparing the predicted using equation (\ref{chance_equation}) and the actual data.}
    \label{fig:f_prod_gr_cts}
\end{figure}

As already discussed, the analyses presented in the previous section do not yield any results unless each outcome has occurred at least once. Therefore, in order to avoid a singularity in the equations (\ref{gen_eqn1}) and (\ref{gen_eqn2}) at $n=0$, calculations are restricted until each outcome occurs at least once. This does not affect the outcomes of the experiment, but presents an initial shift in the cumulatively integrated outcome prediction, as shown in equation (\ref{integration}). Figure \ref{fig:p_and_f_evolution} shows the evolution of the probabilities, outcome frequencies and $\Lambda-$entropy for the coin-toss simulation. Using the evolution equations, a prediction-model is presented in figure \ref{fig:f_H_evols}, which shows prediction of outcome frequencies of \textit{head}, and the growth of uncertainty predicted by equation (\ref{uncertainity_inlimit}). Similarly, the prediction model of the growth rates of \textit{head} outcomes (individually and as a product), and its comparisons to actually observed growth rates are shown in figures \ref{fig:f_gr_ct_S} and \ref{fig:f_prod_gr_cts} respectively.

\begin{figure}
    \centering
   \includegraphics[width=0.8\linewidth]{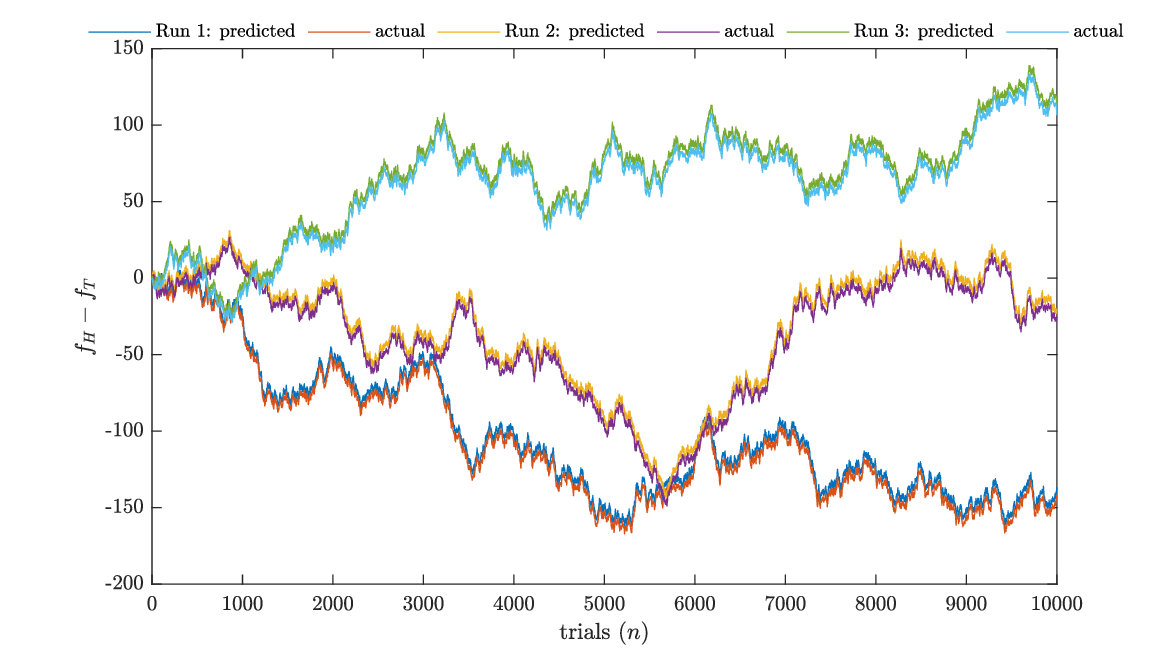}
    \caption{Plot showing different runs of the simulation and the observation for difference in outcome frequencies $f_H-f_T$. It can be seen that as the number of trials grow, the difference becomes more indeterminable for each run. The predicted data and the actual data shown for 3 runs.}
    \label{fig:ft_minus_fh}
\end{figure}

\begin{figure}[!ht]
    \centering
    \includegraphics[width=0.6\linewidth]{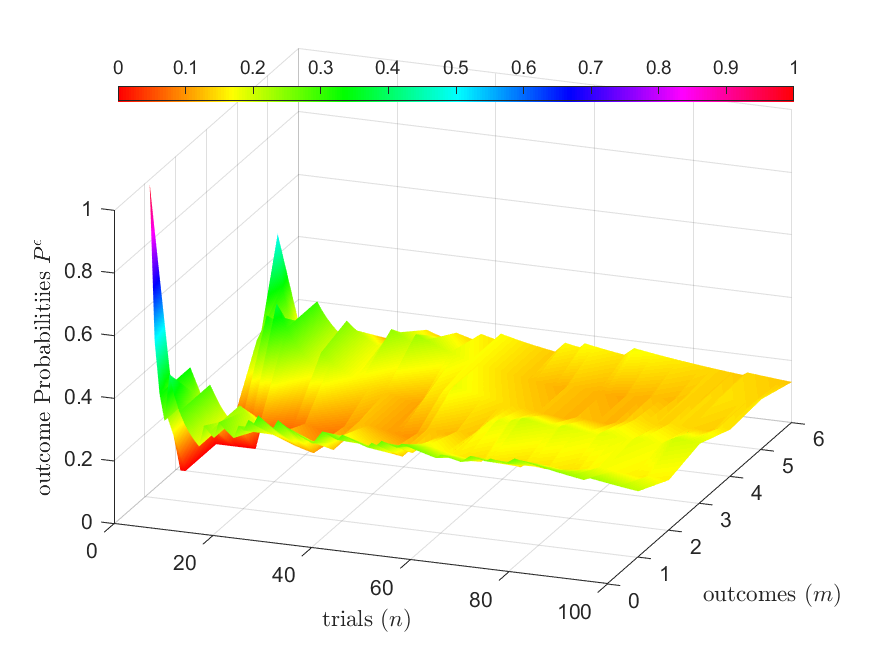}
    \caption{Evolution of empirical probabilities for a coin-toss experiment}
    \label{fig:dice_prob}
\end{figure}

\begin{figure}[!ht]
    \centering
    \includegraphics[width=0.8\linewidth]{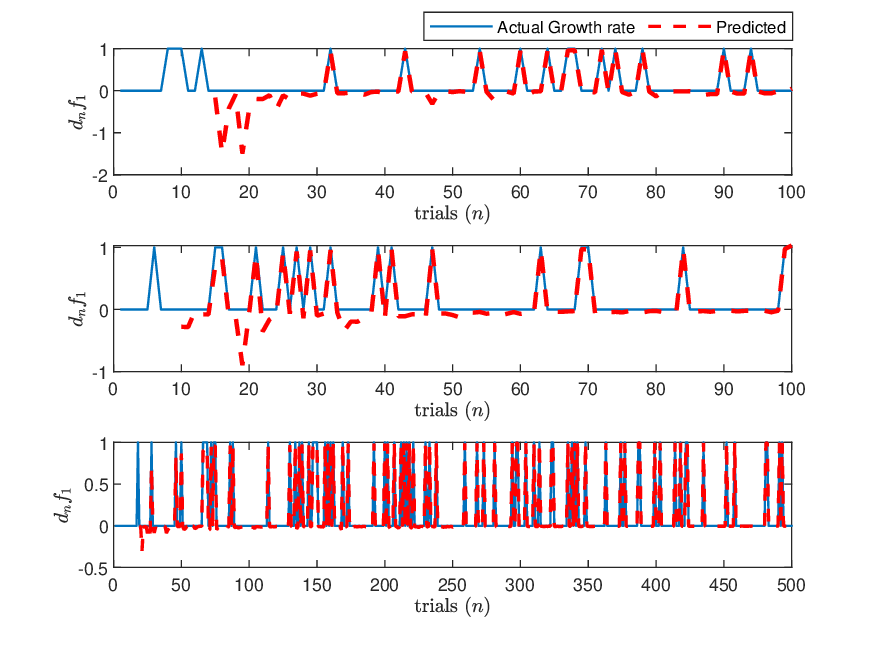}
    \caption{Predicted and actual growth rates of dice outcome as $1$, for different runs. Initial predictions present negative growth rates due to the initial condition, which gets rectified later.}
    \label{fig:dice_rate}
\end{figure}

\begin{figure}[!ht]
    \centering
    \includegraphics[width=1\linewidth]{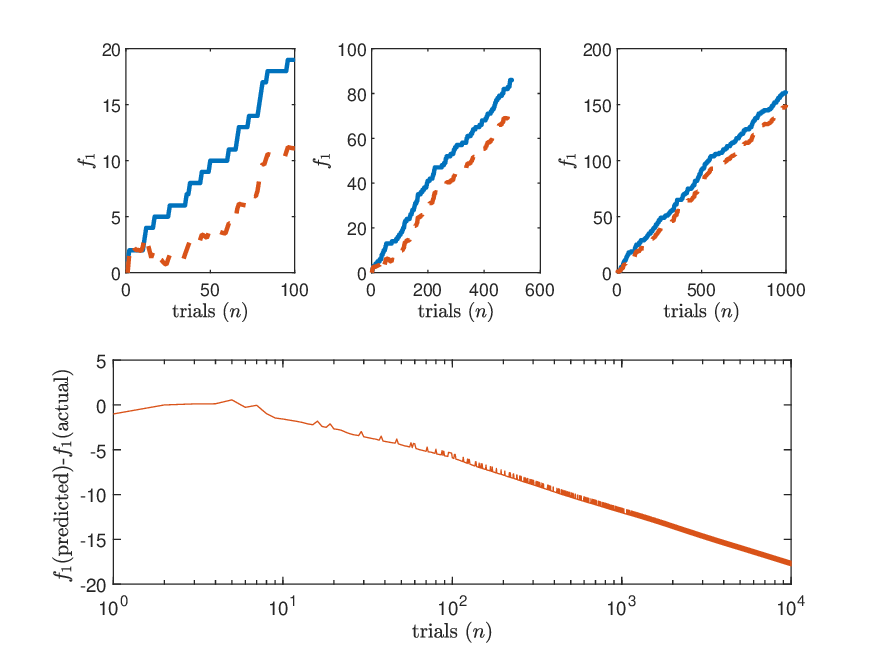}
    \caption{Predicted and actual growth rates of outcome $1$ in the dice experiment. Due to the initial negative growth rate prediction, a deviation in the prediction rates can be seen. Also, the uncertainty in prediction increases due to the growth in number or trials (bottom row).}
    \label{fig:gr_rates_dice}
\end{figure}

\begin{figure}[!ht]
    \centering
    \includegraphics[width=0.8\linewidth]{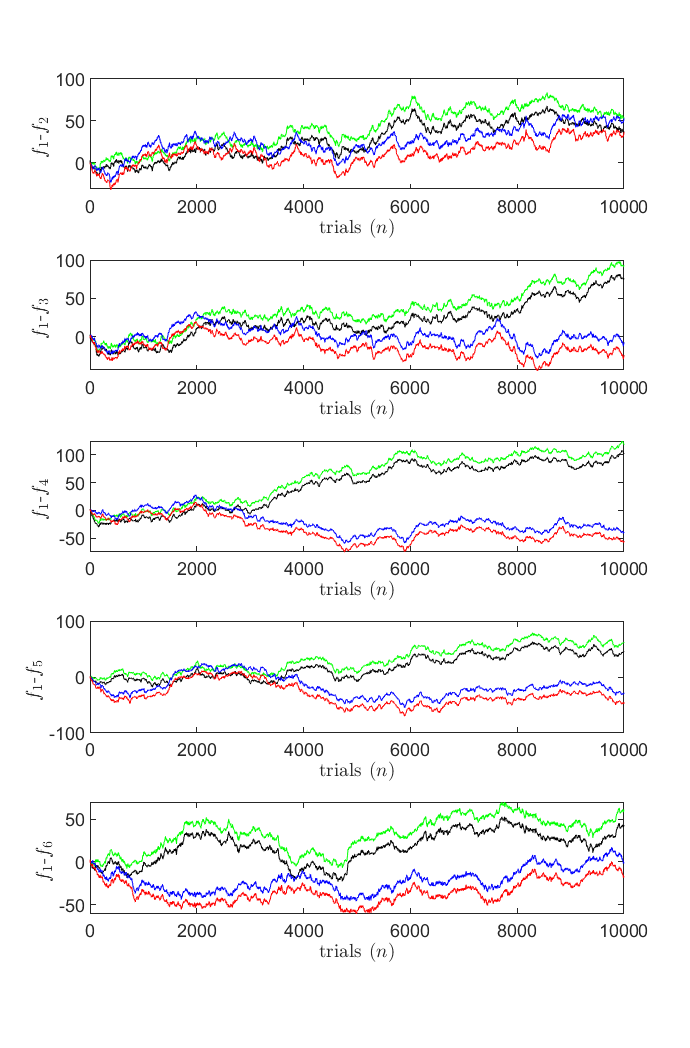}
    \caption{Plots of $f_1-f_i$ for  $i\neq 1$ for the dice roll experiment. Constant offset between predicted (red, black) and empirical (blue, green) values due to initial bias of $f_i (n=0)=1$ for all $i \in [1,6]$.}
    \label{fig:evol_dice}
\end{figure}

Figure \ref{fig:ft_minus_fh} depicts the uncertainty growth in the differences between outcome frequencies of outcomes of the coin-toss simulation, as predicted by equation (\ref{ft-fh}). Upon observing the dice roll experiment, with $m=6$, we see similar predicted results, highlighted in figures \ref{fig:dice_prob}, \ref{fig:dice_rate}, \ref{fig:gr_rates_dice} and \ref{fig:evol_dice}. In the case of the dice experiment, we take initial bias in the experiment with each outcome having initial frequency of occurrence as $1$, due to which a constant offset is observed between the predicted and observed data. This is done in order to deal with the problem of singularities in equation (\ref{gen_eqn1}).

\begin{figure}[!ht]
    \centering
    \includegraphics[width=0.45\linewidth]{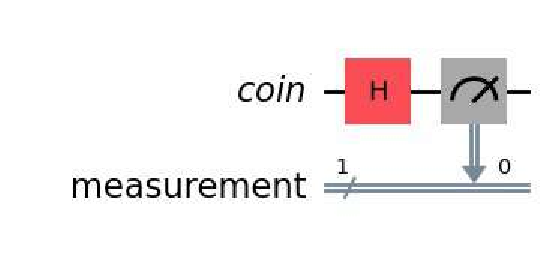}
    \caption{Circuit diagram for the quantum-mechanical experiment.}
    \label{fig:circuit}
\end{figure}

We next perform a quantum-computer experiment in order to test our mathematical analysis. For the same, we use a quantum circuit set-up shown in figure \ref{fig:circuit} in the IBM Quantum-Computer, with the open-user license. We employ a single quantum particle, labelled as the \textit{coin} through a Hadamard gate, after which the \textit{classical register} records the measurement. We expect that over repeated experiments, the number of particles with either \textit{spins} to be $50\%$ each.
\begin{figure}[!ht]
    \centering
    \includegraphics[width=0.45\linewidth]{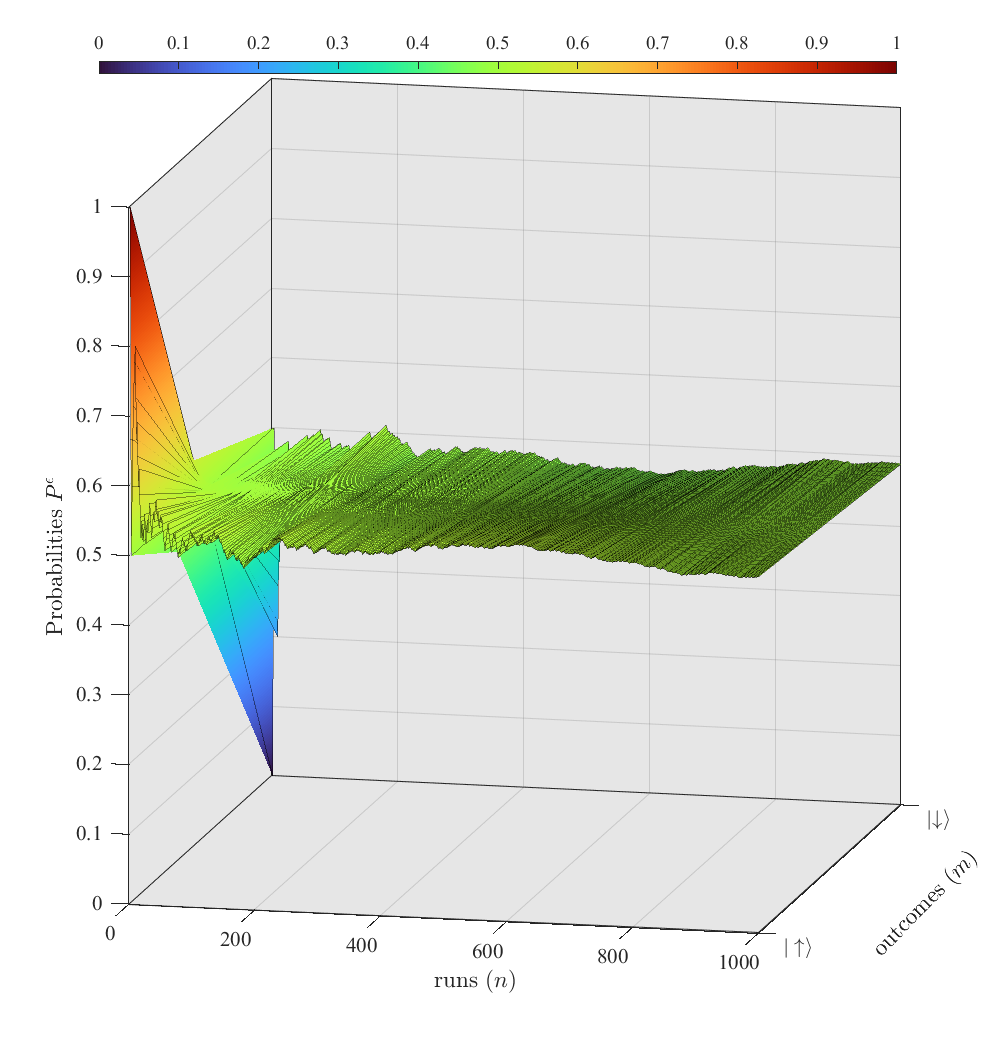}
    \caption{Evolution of probability densities for the quantum experiment.}
    \label{fig:prob_qm}
\end{figure}
\begin{figure}[!ht]
    \centering
    \includegraphics[width=1\linewidth]{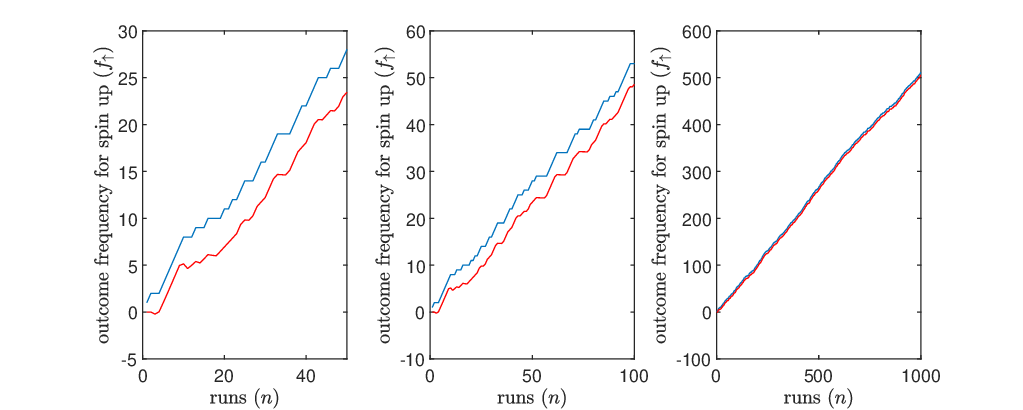}
        \includegraphics[width=0.8\linewidth]{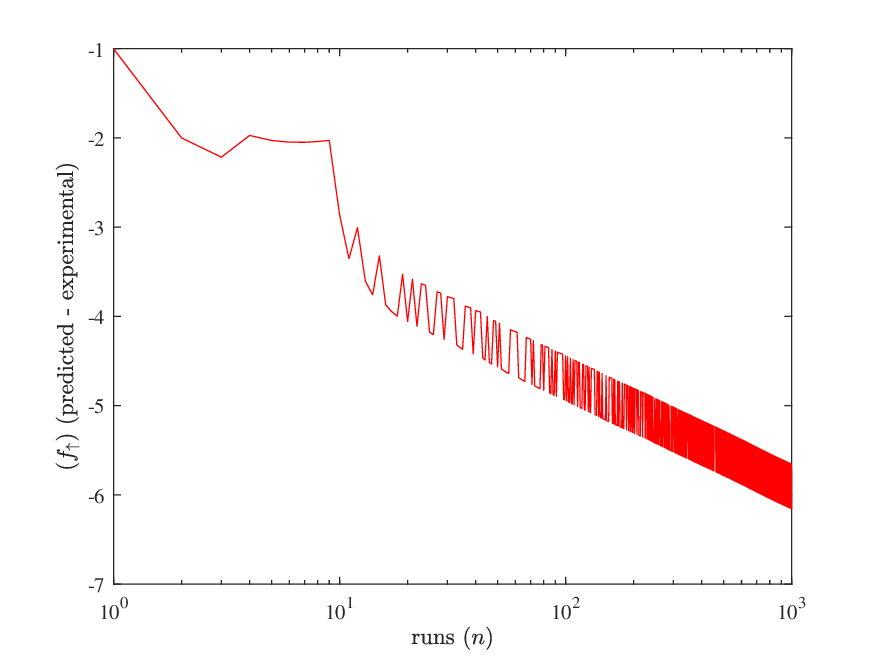}
    \caption{Evolution of the quantum-physical experiment: (above row) Evolution of the outcome frequencies (blue) and the theoretical prediction (red). The initial offset due to restricting the theoretical analysis till each outcome occurs at-least once is visible. (Bottom row) Growth of uncertainty during increase in the number of trials is evident.}
    \label{fig:evol_qm}
\end{figure}

We observe results over $1000$ repeated trials of the experiment. The \texttt{qiskit} code and observed data can be collected from the corresponding author upon request. Figures \ref{fig:prob_qm} and \ref{fig:evol_qm} depict the results and the correspondence with the theoretical predictions.

\section{Dynamical Behaviour and Stability analysis}

Equation (\ref{derivative_of_lambda_entropy_first_principle}) presents a Malthusian-form of $\Lambda-$entropy evolution, with the growth-rate being a function of outcome frequencies $f_i$ and trial number $n$. It can be reframed in the form of a growth equation as follows:
\begin{gather}\allowdisplaybreaks
    \frac{d\Lambda_\sigma}{dn}= h(f,n)\Lambda_\sigma, \quad h(f,n) = \frac{d}{dn}\ln{\prod_{i=1}^n f_i}-\frac{m}{n}.
    \intertext{Expanding for the coin-toss case for simplicity,}
    \frac{d}{dn}P_HP_T = \frac{d}{dn}\left(P_H\cdot(1-P_H)\right) = h(f,n) P_H\cdot(1-P_H)\\
    \Rightarrow \frac{dP_H}{dn}\left(1-2P_H\right) = h(f,n)P_H\cdot(1-P_H),\\
    \Rightarrow \frac{dP_H}{dn} = h(f,n)P_H\frac{1-P_H}{1-2P_H}.\label{nonlinear_eqn_of_probability}
\end{gather}
Equation (\ref{nonlinear_eqn_of_probability}) represents a nonlinear evolution of $P_H$ with a singularity at $P_H=0.5$. In order to test the stability of this evolution, equilibrium points can be found by equating the equation to zero:
\begin{gather}
    h(f,n)P_H\frac{1-P_H}{1-2P_H} = 0 \Rightarrow \begin{cases}
        P_H=0,\\
        P_H=1,\\
        h(f,n)=0.
    \end{cases}
    \intertext{The third case, $h(f,n)=0$ implies the condition when }
    \frac{d}{dn}\ln{\prod_{i=1}^n f_i}=\frac{m}{n}\Rightarrow \ln{\prod_{i=1}^n f_i} = m\ln{n} + C.
    \intertext{It can be determined that this condition applies when the ratio $\prod_{i=1}^m f/n^m\rightarrow$ constant, which occurs when $n\rightarrow \infty$. The two equilibrium points $P_H^*$ therefore occur at $P_H^*=\{0,1\}$. The stability of these equilibria points can be analysed based on their growth rates with respect to $P_H$:}
    h(f,n)\frac{\partial}{\partial P_H}\left(\frac{P_H(1-P_H)}{1-2P_H}\right)=h(f,n)\frac{1+2P_H^2 - 2P_H }{(1-2P_H)^2} = \begin{cases}
        h(f,n) & \text{at }P_H^*=0,\\
        h(f,n) & \text{at }P_H^*=1.
    \end{cases}\label{stability}
\end{gather}
It can be seen from the presented analysis in equation (\ref{stability}) that the stability of the evolution of probability $P_H$ depends on the growth rate $h(f,n)$. Therefore, for negative values of $h(f,n)$, the probability $P_H$ decay either towards $0$ or $1$. For values of $h(f,n)$ greater than zero, the probability $P_H$ tends towards the singularity $0.5$ but never reaches the value. Perhaps, this is the true reason behind Bernoulli's Law of Large Numbers. As the experiment repeats, the probability growth rate $h(f,n)$ grows from negative to positive values. This causes the probability to deviate from the stable points $0$ and $1$. The instability causes the probability to transition between the two stable points, either from $0$ to $1$ or in the opposite direction. However, the singularity present at $0.5$ causes the value of $P_H$ to tend to increase towards $0.5$ as $n\rightarrow\infty$. Since the probability never actually reaches $0.5$, it shows an asymptotic behaviour and almost becomes a constant, and therefore, becomes equal to the theoretically predicted value $P_H^0 = 0.5$. We now proceed to the concluding remarks of our work in the next section. 


\section{Conclusions}
In this work, we have extended the mathematical analyses and interpretations of Bernoulli's Golden theorem by presenting some interesting analytical predictions. The empirical probabilities of random outcomes, the nature, randomness and predictability of outcomes in a random experiment and the mathematics depicting the meaning of \textit{chance}, along-with its influence on outcomes of a repeated random experiment have been shown in this work. 
The analyses presented in this work, apart form depicting the nature of flow of random experiments in repetitions, also present a technique of measuring the empirical randomness of a dataset. The dynamical evolution analyses presented in this work showcases the analytical as well as observational intricacies of the \textit{Golden theorem}, including why it occurs. It presents a qualitative relationship between a nonlinear divergence and the idea of empirical randomness of an otherwise random experiment. It also presents an appreciation of the beauty of Bernoulli's work and its applications by extension.
\bibliography{references}

\begin{thebibliography}{10}

\bibitem{Bellhouse2005DecodingAleae}
D.~Bellhouse, ``{Decoding Cardano's Liber de Ludo Aleae},'' {\em Historia Mathematica}, vol.~32, no.~2, 2005.

\bibitem{Mattmuller2014The1713}
M.~Mattm{\"{u}}ller, ``{The difficult birth of stochastics: Jacob Bernoulli's Ars Conjectandi (1713)},'' {\em Historia Mathematica}, vol.~41, no.~3, 2014.

\bibitem{Bolthausen2013BernoullisNumbers}
E.~Bolthausen and M.~V. W{\"{u}}thrich, ``{Bernoulli's law of large numbers},'' 2013.

\bibitem{Chibisov2016BernoullisNumbers}
D.~M. Chibisov, ``{Bernoulli’s law of large numbers and the strong law of large numbers},'' {\em Theory of Probability and its Applications}, vol.~60, no.~2, 2016.

\bibitem{Teran2008OnNumbers}
P.~Ter{\'{a}}n, ``{On convergence in necessity and its laws of large numbers},'' {\em Advances in Soft Computing}, vol.~48, 2008.

\bibitem{Weba2009AMethod}
M.~Weba, ``{A quantitative weak law of large numbers and its application to the delta method},'' {\em Mathematical Methods of Statistics}, vol.~18, no.~1, 2009.

\bibitem{Goldstein1975SomeNumbers}
J.~A. Goldstein, ``{Some applications of the law of large numbers},'' {\em Boletim da Sociedade Brasileira de Matem{\'{a}}tica}, vol.~6, no.~1, 1975.

\bibitem{Dedecker2007ApplicationsNumbers}
J.~Dedecker, P.~Doukhan, G.~Lang, L.~R. Jos{\'{e}}~Rafael, S.~Louhichi, and C.~Prieur, ``{Applications of strong laws of large numbers},'' in {\em Weak Dependence: With Examples and Applications}, 2007.

\bibitem{Teran2006AApplications}
P.~Ter{\'{a}}n and I.~Molchanov, ``{A general law of large numbers, with applications},'' {\em Advances in Soft Computing}, vol.~37, 2006.

\bibitem{SHIRIKYAN2003AAPPLICATIONS}
A.~SHIRIKYAN, ``{A VERSION OF THE LAW OF LARGE NUMBERS AND APPLICATIONS},'' 2003.

\bibitem{Yang2008AApplications}
S.~Yang, C.~Su, and K.~Yu, ``{A general method to the strong law of large numbers and its applications},'' {\em Statistics and Probability Letters}, vol.~78, no.~6, 2008.

\bibitem{Kay2014LawsApplications}
J.~W. Kay, ``{Laws of Large Numbers: Applications},'' in {\em Wiley StatsRef: Statistics Reference Online}, 2014.

\end{thebibliography}
\bibliographystyle{ieeetr}
\section*{Miscellaneous Information}
\textbf{Author Contributions:} Conceptualization, A.L.; methodology, A.L. and S.A.; software, A.L. and S.A.; validation, A.L. and S.A.; formal analysis, A.L.; investigation, A.L. and s.A.; resources, A.L. and S.A.; data curation, A.L. and S.A.; writing---original draft preparation, A.L.; writing---review and editing, S.A.; visualization, A.L. and S.A.; All authors have read and agreed to the published version of the manuscript.
\\

\noindent\textbf{Funding:} This research received no external funding.\\

\noindent\textbf{Data availability:} Numerical Data and associated codes are available with the corresponding author (A.L. and S.A.) upon request. \\

\noindent\textbf{Acknowledgments:} One of the authors (A.L.) thanks A. Thapa for motivation and discussions. Both authors acknowledge and appreciate the availability of IBM Quantum computing facility for free, which enabled some analyses reported in this work.\\

\noindent\textbf{Conflicts of interest:} The authors declare no conflicts of interest.

\end{document}